\documentclass[preprint,aps,prb,final,amsfonts,amssymb,nobibnotes,amsmath,floatfix,showpacs]{revtex4}
\usepackage{graphicx}

\begin{document}

\title{Electronic structure and Peierls instability in graphene nanoribbons sculpted in graphane}

\author{Valentina Tozzini}
\email{tozzini@nest.sns.it}
\affiliation{Scuola Normale Superiore, NEST CNR-INFM and IIT-Italian Institute of Technology, I-56100 Pisa, Italy}

\author{Vittorio Pellegrini}
\affiliation{Scuola Normale Superiore, NEST CNR-INFM and IIT-Italian Institute of Technology, I-56100 Pisa, Italy}



\date{\today}

\begin{abstract}

Graphene nanoribbons are semiconductor nanostructures with great potentials in nanoelectronics. Their realization particularly with small lateral dimensions below a few nanometers, however, remains challenging. Here we theoretically analyze zig-zag graphene nanoribbons created in a graphane substrate (a fully saturated two-dimensional hydrocarbon with formula CH) and predict that they are stable down to the limit of a single carbon chain. We exploit density functional theory with B3LYP functional that accurately treats exchange and correlation effects and demonstrate that at small widths below a few chains these zig-zag nanoribbons are semiconducting due to the Peierls instability similar to the case of polyacetylene. Graphene nanoribbons in graphane might represent a viable strategy for the realization of ultra-narrow semiconducting graphene nanoribbons with regular edges and controlled chemical termination and open the way for the exploration of the competition between Peierls distortion and spin effects in artificial one-dimensional carbon structures. 

\end{abstract}
\pacs{    }

\maketitle 

Graphene is a single atomic layer of carbon atoms arranged in a honeycomb lattice.  This system is remarkably appealing both for fundamental studies and potential electronic applications since it hosts a high-mobility two-dimensional (2D) electron or hole gas displaying unique quantum transport properties  \cite{nov04,geim07,zheng05}. The unconventional properties of graphene are the result of its peculiar band structure having zero gap and linear dispersion of conduction and valence bands at the corners of the Brillouin zone (Dirac points). Graphene-based nanoelectronic logic devices require, however, the development of graphene nanostructures displaying sufficiently large band gaps. This is currently achieved in narrow graphene ribbons (graphene nanoribbons or GNR), which represent so far one of the most promising strategies for graphene nanoelectronics \cite{berger06,li08}.  GNRs can be nowadays fabricated by exploiting different chemical, or lithographic methods \cite{li08,tap08,han07}. Although GNR field-effect transistors have been demonstrated \cite{li08,wang08},  and much effort is being devoted to GNR fabrication, the realization of GNRs with controllable and reproducible properties remains a challenge. This is mainly caused by the roughness present at the physical edges of the nanoribbon and by the large dependence of the electronic properties on the chemical edge termination and on the nanoribbon chirality both of which cannot be finely controlled in the employed methods of fabrication \cite{han07,lu09,lee09,jia09}.  In addition, current approaches do not allow to reach ultra-narrow GNRs with widths of just one or a few chains but are limited to values of the order of 3nm (N$\sim$15 chains) or above. 
\par
Here we analyze a different type of GNRs that is obtained by ÒsculpturingÓ graphane substrates. We recall that graphane, a fully saturated hydrocarbon version of graphene, is obtained by adding hydrogen atoms to graphene with stoichiometry 1:1\cite{sofo07}. Its structure, stability and electronic properties were first theoretically predicted for symmetric configurations where hydrogen atoms are bound half on one side and half on the other side of the graphene sheet, and recently experimentally observed \cite{elias09}. Graphane is an insulator, with a band gap of $\sim$3.5eV, and its most stable conformation has the same in-plane symmetry of graphene and their lattice parameters differ only by 4\% \cite{sofo07}. This circumstance suggests that hybrid stable graphane/graphene nanostructures with peculiar electronic properties could be built with a designed shape by selectively removing hydrogen atoms at specific locations. In particular GNRs defined in graphane  (Graphane/Graphene nanoribbons or GGNRs) as shown in the cartoons of Fig.1 can have in principle lateral atomic dimension down to the single chain and well-defined chirality if, for example, an STM is used for nanoscale pattering by removal of the hydrogen atoms \cite{shen95,lyding94}. Other methods to remove hydrogen atoms could be envisioned such as laser heating or e-beam irradiation. 

In order to explore the reliability of this idea we theoretically analyze the stability and the structural and electronic properties of zig-zag GGNRs (elongated along the y direction as shown in Fig. 1) within the density functional theory (DFT) frame, including electronic exchange and correlation at different levels of accuracy. We address the stability of such nanoribbons and demonstrate they are semiconducting even in the ultra-narrow limit of a single carbon chain. We show that the opening of the gap at very small widths is the result of the Peierls distortion that leads to bond-length alternation (difference in lengths of subsequent bonds or BLA) along y direction and removal of the degeneracy between the highest occupied state (HOMO) and the lowest empty state (LUMO) as in the case of the polyacetylene \cite{longuet59}, which has the same structure of the single chain zig-zag GNR. We show that, as for polyacetylene, the Peierls instability is captured only including the appropriate form of the exchange-correlation energy functional. The results predict a band-gap tunability of such GGNRs in the range between 0.2 eV and 1.5 eV, with the latter value obtained in the limit of a single chain with atomic lateral dimension. In the case of GNRs, the band-gap dependence on the width is found in excellent agreement with available experimental data. These results demonstrate that GGNRs could be exploited as building blocks for the realization of complex electronic circuits directly sculptured in the graphene substrate and for the exploration of the impact of Peierls instability in carbon nanostructures.

The DFT calculations here performed are based on both the extended (plane waves, PW) and the localized (Gaussian bases, GB) wavefunction approaches. The first method has better performances on extended systems and is therefore used for the dynamical calculations, while the second should better reproduce the structural properties of confined systems and is used for the study of the electronic structure and geometry optimization. Within the PW scheme the 1s core electrons for C atoms were implicitly treated with TroullierÐMartins pseudopotentials, and the valence electrons wavefunctions were expanded in plane waves with an energy cutoff of 70 ryd (90 Ryd in selected cases). Supercell including up to six unitary cells in the y direction were used. These calculations were performed with the CPMD3.13 code \cite{CPMD}. Within the GB scheme all-electron calculations were performed using the 6-31G$^\ast$ basis set. The minimal unitary cell was sampled with up to 200 k points. The GB calculations were performed with Gaussian03 \cite{frisch04}.  In both cases different energy functional were used: the BeckeÐLeeÐYangÐParr (BLYP) \cite{becke88} exchange and correlation functional, its hybrid version (B3LYP) including 20\% of explicit exchange \cite{becke93}, and the LSDA \cite{ceperley80,vosko80},  functional explicitly treating the spin density. The starting graphane configuration was build using the structural parameters given in Ref.[12] and subsequently optimized (both structure and cell parameters) with standard local minima search algorithms. In each case the graphene ribbons were obtained by simply removing the hydrogen atoms in specific locations as shown in Fig 1 and re-optimizing the structure. The molecular dynamics simulations were performed using the BLYP functional within the Car-Parrinello approach \cite{car85} using the electronic mass preconditioning scheme \cite{tassone94} and time-step of 0.193fs. We also studied GNRs with the aim of comparing their properties with those of corresponding GGNRs. In the case of GNRs, the edges are saturated with hydrogen, a natural termination that does not introduce states within the gap. The GGNRs are even more naturally terminated, each dangling C- bond being homogeneously saturated with C, although the graphane introduces some strain due to the 4\%  
lattice parameter mismatch between graphane and graphene. This, however, does not significantly contribute to modify the electronic/structural properties of GGNRs with respect to the saturated GNRs. 
Figure 2a shows the optimized structure of the zig-zag of both GNRs and GGNRs. In the case of GNRs, the C-C bonds display a marked BLA along the x direction that rapidly decreases (within $\sim$3 chain from the edge) merging into the graphene-like configuration as one moves towards the center of the ribbon. A similar behavior is seen for the ribbons embedded in graphane, although the strain induced by the graphane lattice imposes slightly different C-C values. In addition, a similar decaying C-C BLA can be seen within the graphane moving out from the nanoribbon. In order to address the stability of the GGNRs, we first heated the system up to $T <600-700$K and observed no hydrogen hopping or other substantial distortions other than due to the heating itself (data not shown). This indicates that once formed by selectively removing the hydrogen atoms, the GGNRs are robust with respect to hopping processes of hydrogen atoms from the graphane matrix into the graphene ribbon.
In order to evaluate the energy barrier for such a process, we performed constrained molecular dynamics on the reaction path illustrated in the lower panel of Fig.2b. The hydrogen atom highlighted in red is forced to hop to its nearest neighbor, without the possibility of going back. From this unstable configuration the atom spontaneously moves to its final position yielding a configuration, representing a terminated wire embedded in graphane, chemically consistent (i.e. no radical involved as in the intermediate) but nevertheless less stable than the starting one of $\sim$1.5 eV. The activation barrier is of $\sim$5 eV. The upper panel of Fig.2b reports the evolution of the three involved C-H distances along the reaction coordinate.

Having shown that GGNRs once formed are stable we now discuss their electronic properties. The calculated gap using the B3LYP functional for both H-passivated GNRs (filled red dots) and GGNRs (filled blue dots) is shown in the main panel of Fig.4 as a function of the wire width W. This approach neglects the spin polarization that is taken into account in the LSDA calculations shown as squares (empty squares are data from Ref. [\cite{son061}]). The data in Fig.4 demonstrate that the spin-induced opening of the gap vanishes at small widths in agreement with previous calculations \cite{son061,yang07,son06}.  It should be noted, in addition, that the curve extrapolated from the experimental data obtained in large nanoribbons (red line) naturally merges with the experimental value of the gap of the single-chain nanoribbon, i.e. the polyacetylene (red asterisk). This suggests significant energy gaps of around $0.5-1.5$ eV at nanoribbon widths where spin effects are negligible. These large gaps are indeed obtained by our DFT analysis (red and blue filled dots). In this small-width regime the HOMO-LUMO degeneracy is removed by the Peierls distortion that leads to BLA along the y direction (in addition to BLA along the x direction) as shown in the inset to Fig.4 in agreement with recent calculation based on a similar approach \cite{pisani07}.  We found that the Peierls distortion is the dominant mechanism responsible for the opening of the gap in both GNRs and GGNRs up to a number of chain $N=3-4$. For GGNRs the slightly different values of gap and BLA compared to those in GNRs are due to the small strain induced by the graphane matrix. For $N=4$ the BLA value is very small, although it results still in non-negligible values of the gap. For $N>4$ the spin mechanism becomes important. 

In order to highlight the impact of the different contributions to the band-gap, we report in Fig. 4 the schematic structures of the $\pi$ and $\pi^\ast$ bands in zig-zag nanoribbons with variable number N of chains as the various effects described above are included starting from the ideal hexagonal geometry (all equal C-C bonds). Each added chain adds a couple of $\pi?\pi^\ast$ bands. The edge states start to appear at a certain width as an effect of the accumulation of degenerate states between $2/3\pi/a<k<\pi/a$. When the geometry relaxation is allowed, even at the lower-level theory, the BLA effect in the x direction is observed, similar to the reconstruction of the surfaces of a 3D crystal. This modifies the bands or breaks the symmetry mainly at the $k=\pi/a$ point, but the zig-zag nanoribbon remains metallic at low values of the width as in Refs.\cite{son061,yang07,son06,barone06}. When the B3LYP functional with explicit exchange is used, the gap opening at $\pi/a$ at small nanoribbon width (low values of N) due to Peierls distortion (and BLA in y direction) can be seen. The effect is still appreciable up to N=4, where the gap opening occurs in one or more points between $2/3\cdot\pi/a$ and $\pi/a$. When the spin-dependent functional is considered, the edge states degeneracy is removed. The effect is absent for N=1 and N=2 where there are no edge states, and rapidly decrease as N increases because the edges are less and less interacting. This behavior is also in agreement with previous calculations \cite{son061,yang07,son06}. 

In summary, we have shown that graphene nanoribbons sculpted in graphane by selectively removing hydrogen atoms display the stability and semiconducting properties required for nanoelectronic applications even when their lateral dimension is within the atomic limit. Such approach of removing hydrogen atoms from graphane might offer a promising strategy for the realization of ultra-narrow nanoribbons with ideal edges.


\begin{acknowledgments}
We thank Fabio Beltram, Giuseppe Grosso, Giuseppe Pastori Parravicini and Paolo Giannozzi for useful discussions. We aknowledge the allocation of computer resources on CINECA national facility by means of INFM-CNR ÒProgetto di Calcolo Parallelo 2008-2009Ó.
\end{acknowledgments}


\newpage

\begin{figure}
\begin{center}
\includegraphics*[width=12cm]{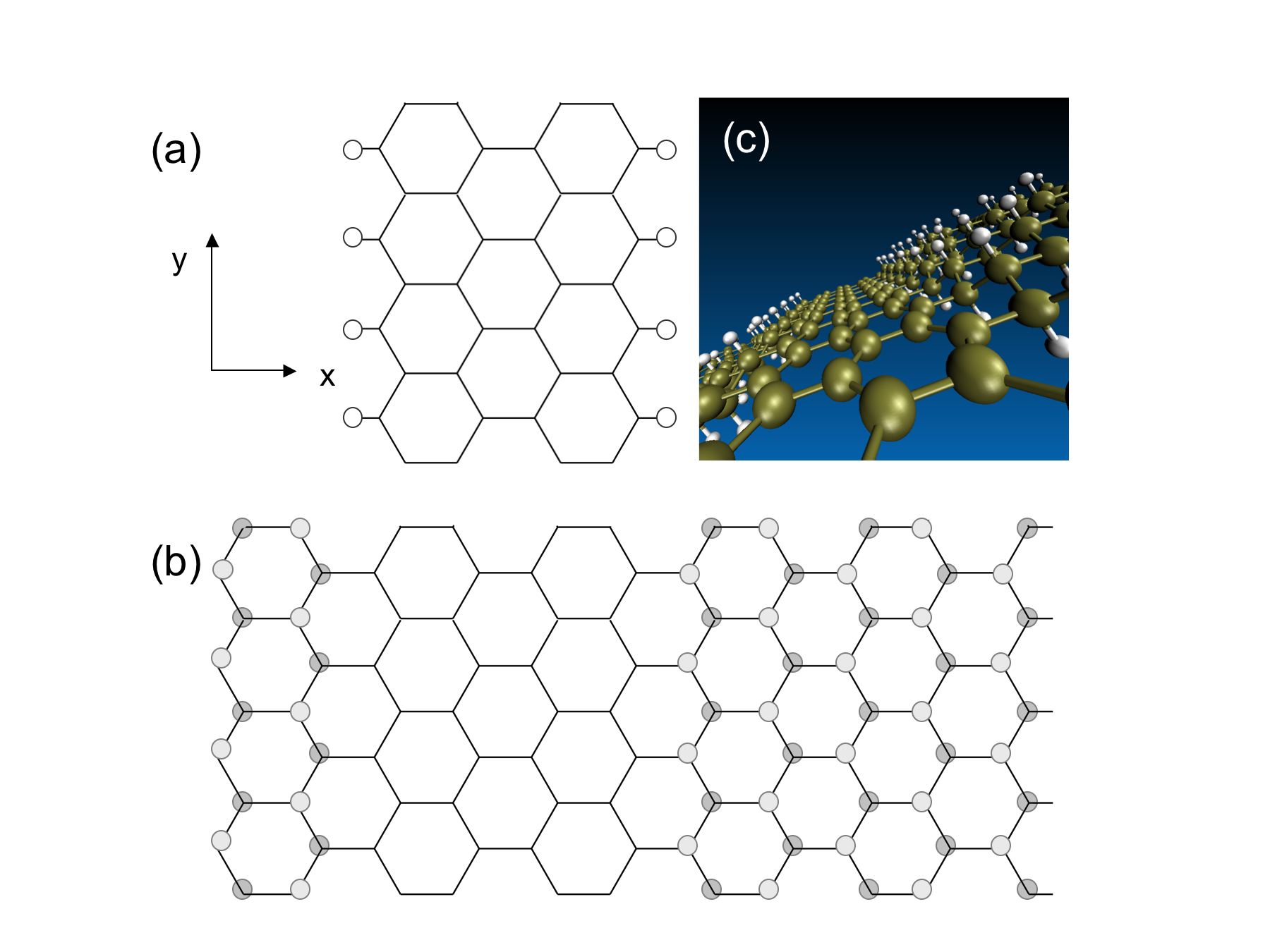}
\end{center}
\caption{(a) Schematic representation of a graphene nanoribbon (GNR) composed by four zig-zag chains passivated with H. White balls represent H atoms, coplanar with C atoms located at the vertices of the honeycomb lattice. The system is periodic in the y direction. (b) Schematic representation of a graphane/graphene nanoribbon (GGNR) composed by four chains embedded in graphane. Light grey and dark grey balls represent H atoms bonded to the C atoms of the corresponding honeycomb lattice site and located above and below the honeycomb lattice plane, respectively. (c) A perspective view of a GGNR with 3 chains. C and H atoms are represented as brown and white balls, respectively. }
\end{figure}
\begin{figure}
\begin{center}
\begin{tabular}{cc}
\includegraphics*[height=8cm]{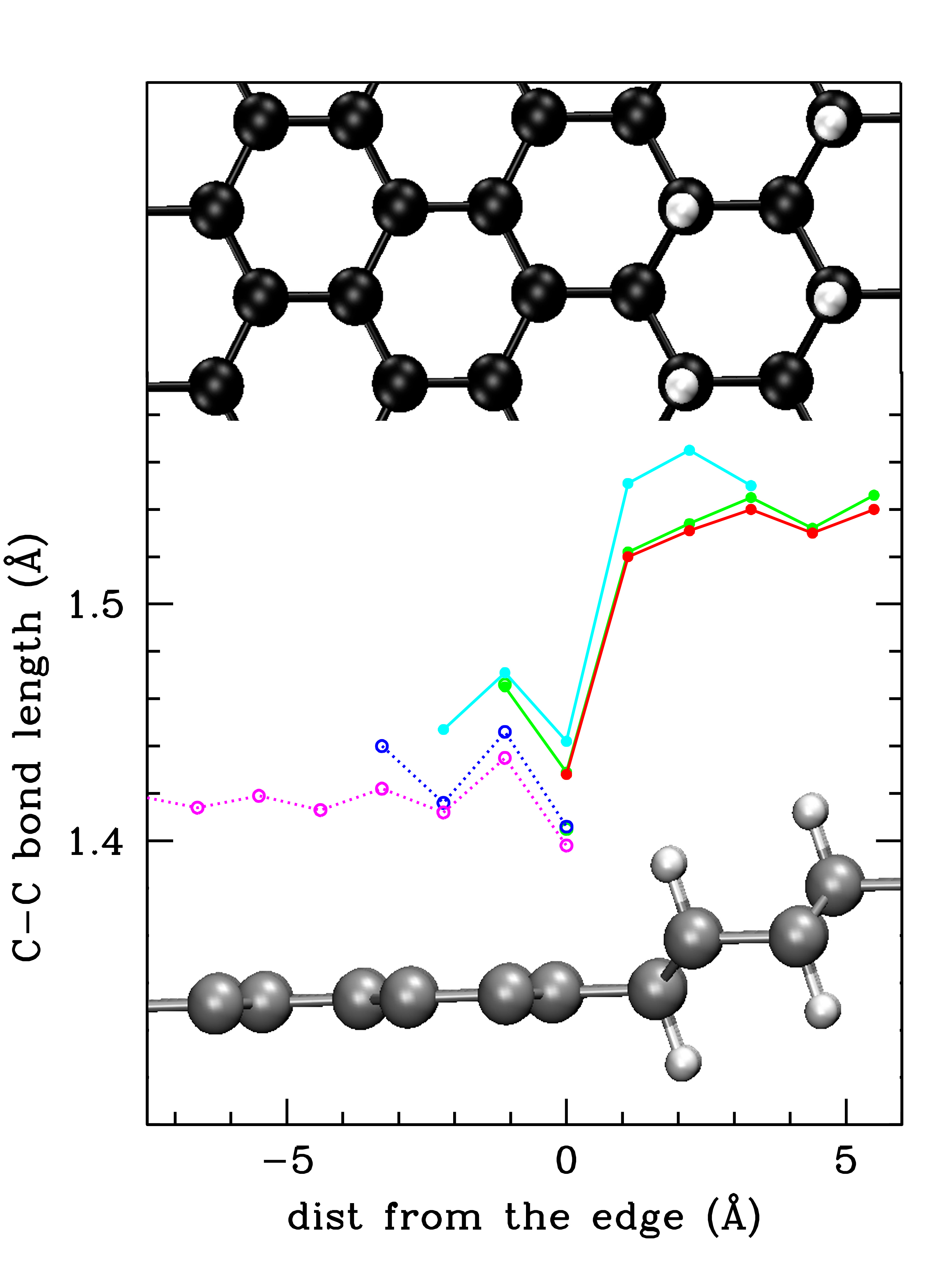} & \includegraphics*[height=8cm]{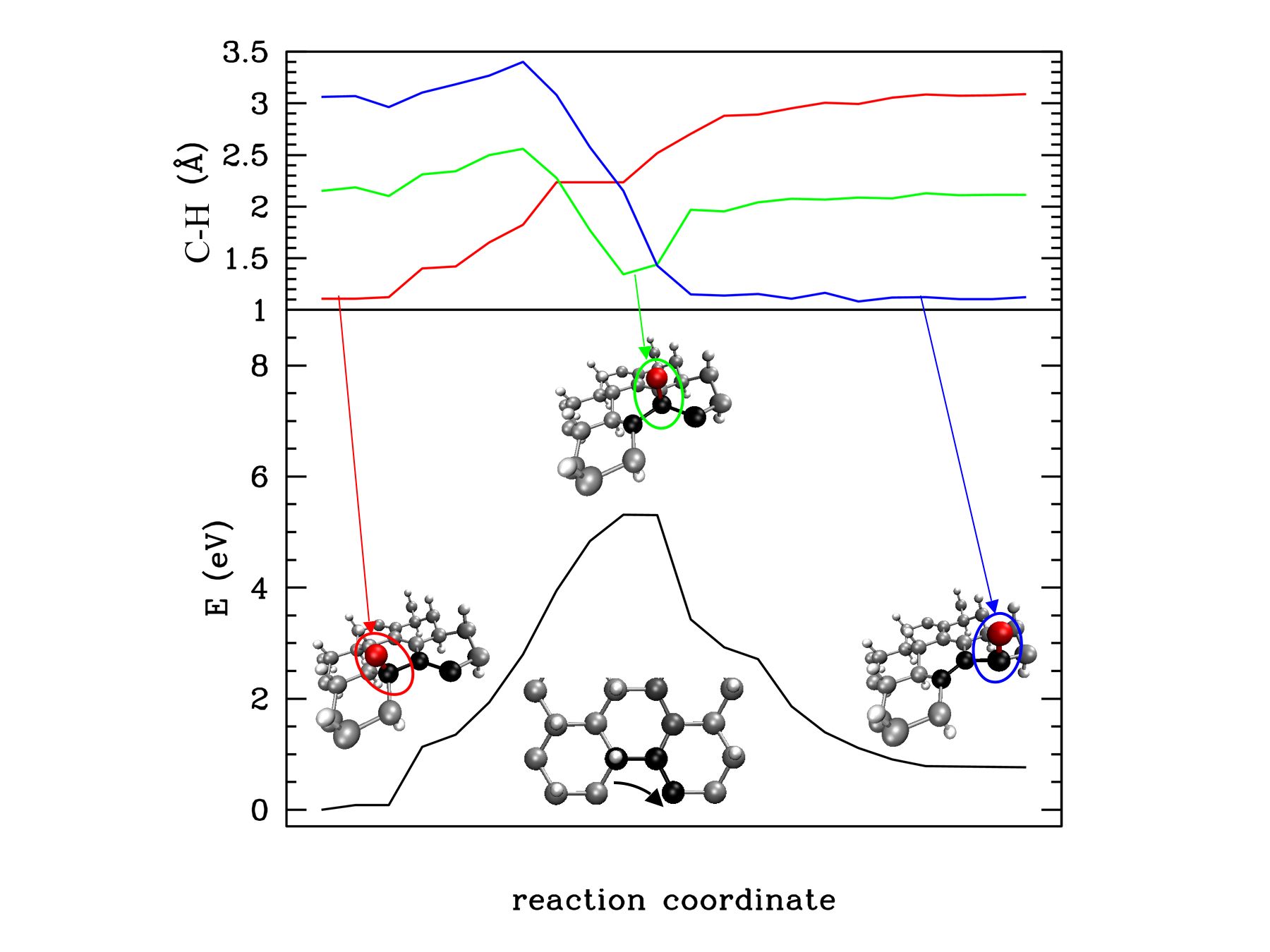}\\
(a)&(b)\\
\end{tabular}
\end{center}
\caption{a) The cartoons correspond to the optimized structure of the zig-zag graphene/graphane nanoribbon (GGNRs). Horizontal axis corresponds to the x direction as defined in Fig.1. The variation of the C-C bond length along the x direction is also reported (origin at the wire edge, negative distances are within the ribbon, from the edge to the ribbon center; positive distances are within the graphane matrix; when BLA is present in the orthogonal direction, the points correspond to the the average value of the C-C bond length). Solid lines, filled dots refer to ribbons embedded in graphane; dotted lines, empty dots refer to H-passivated nanoribbons. Red, green, cyan, blue and magenta correspond to the results of zig-zag GGNRs with widths of 1, 2, 3, 4, and 8 chains, respectively. (b) Simulation of the hopping process of a hydrogen atom from the graphane substrate to the graphene chain. The black line shows the system energy profile along the reaction path defined by the three black carbon atoms in the cartoon at the bottom of the figure. As shown in the three cartoons placed along the energy profile, the hopping hydrogen (in red) is forced to leave the graphane and hop to the graphene wire passing from a intermediate site. The variation of the three C-H bond distances along the path is also shown in the upper part of the figure, in red green and blue. }
\end{figure}

\begin{figure}
\begin{center}
\includegraphics*[width=12cm]{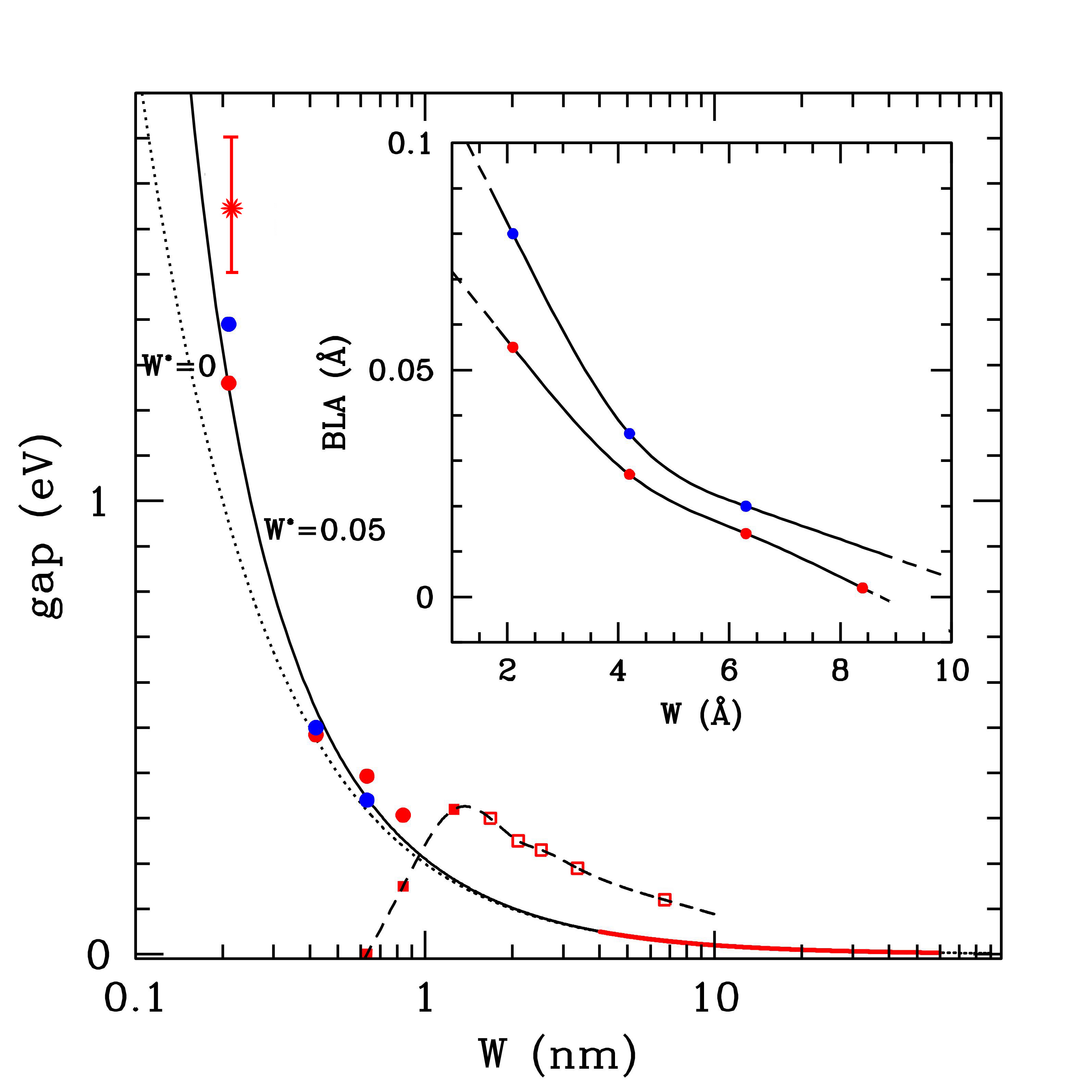}
\end{center}
\caption{Main panel: energy gap versus wire width (W). Dots and squares: results of calculations, red for H-passivated zig-zag graphene nanoribbon (GNR), blue for zig-zag graphene/graphane nanoribbon (GGNR). Filled circles: B3LYP calculations with no spin polarization. Squares: LSDA calculations with spin polarization (empty squares are data from Ref. [21]; the dashed line is a guide to the eye). Solid and dotted lines display the function gap=A/(W-W$^\ast$) with A=0.2 eVnm, fitted from experimental data (Ref.[7]) and extrapolated at small W (the experimental data range is reported as a red band on the curve). Red asterisk with error bar corresponds to the experimental gap of polyacetylene. Inset: Bond Length Alternation (BLA) along the y axis (see Fig.1) versus the nanoribbon width. Colors and symbols as in the main panel. The lines are guides to the eye.
}
\end{figure}
\begin{figure}
\begin{center}
\includegraphics*[width=12cm]{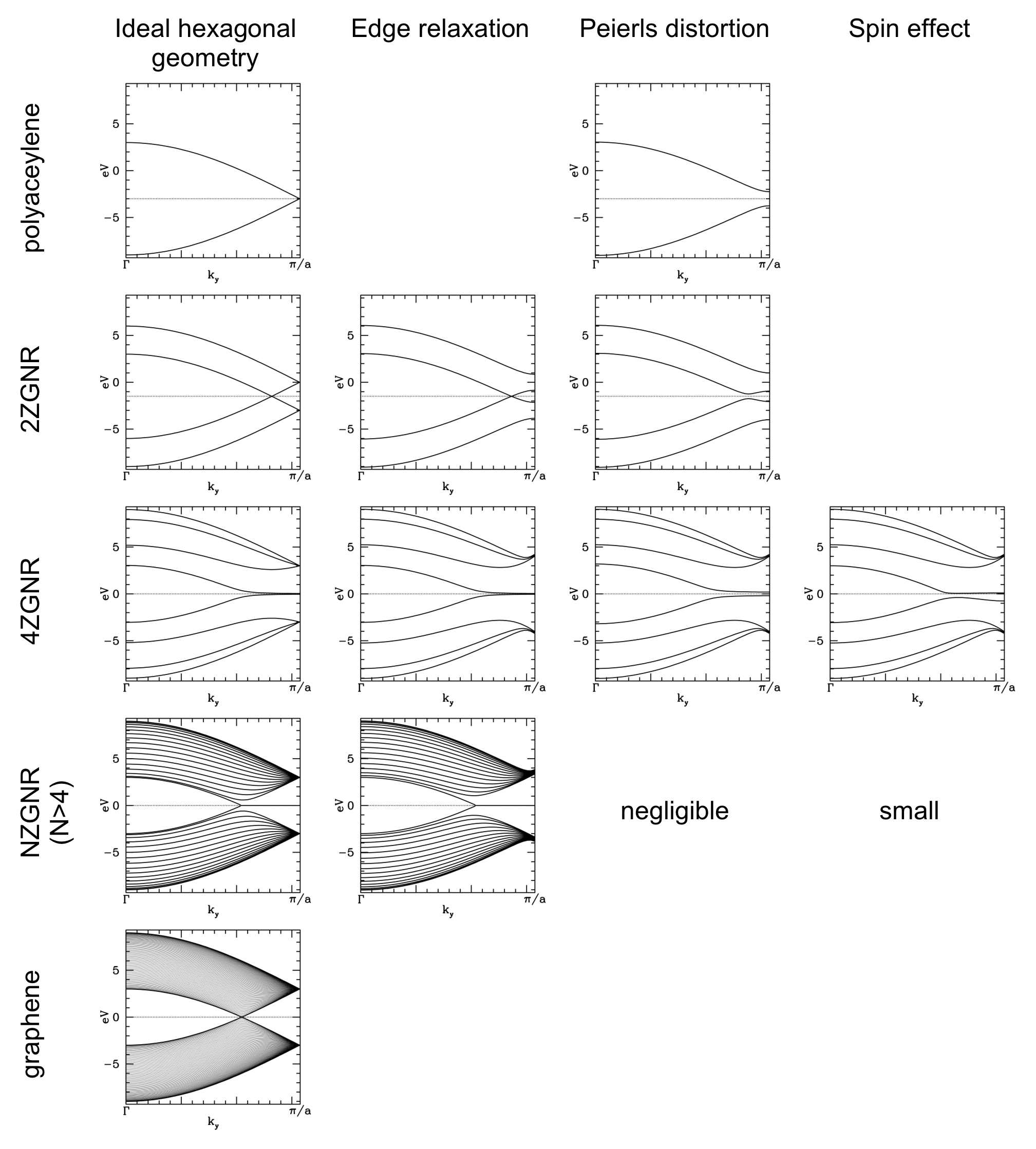}
\end{center}
\caption{Schematic qualitative representation of $\pi$ and $\pi^\ast$ band structures in graphene zig-zag nanoribbons of different widths (N is the number of chains) and at different levels of sophistication of the theoretical analysis: first column corresponds to the ideal hexagonal geometry. $k=2/3\pi/a$ corresponds to the Dirac K point of graphene Brillouin zone, where in fact the HOMO and LUMO bands converge in the limit of large N. The second column shows the results in the case of geometric  }
\end{figure}
\clearpage
\newpage

\end{document}